\documentclass[conference, onecolumn, a4paper, 11pt]{IEEEtran}      
\pdfoutput=1
\usepackage[utf8]{inputenc}
\usepackage{amsmath,amsfonts,amsbsy,amssymb}
\usepackage{amssymb,amsthm,mathrsfs,amsfonts,dsfont} 
\usepackage{mathabx}
\usepackage{mathrsfs}
\usepackage{tabularx}
\usepackage{graphicx}
\usepackage{cite}
\usepackage{wasysym}
\usepackage{multirow}
\usepackage{float}
\usepackage{color}
\usepackage{epstopdf}
\usepackage{commath}
\usepackage{upgreek}
\usepackage{ textcomp }

\usepackage{graphics} 
\usepackage{graphicx}
\usepackage{siunitx}
\usepackage[nolist]{acronym}
\usepackage{mathtools}

\usepackage{xcolor}

\newcommand{\matriz}[1]{\textbf{#1}}

\title{Reconfigurable Intelligent Surface-Aided Grant-Free Access for Uplink URLLC}

\author{Dick Carrillo Melgarejo$^{1}$, Charalampos Kalalas$^{2}$, Arthur Sousa de Sena$^{1}$,  Pedro H. J. Nardelli$^{1}$\\ , and Gustavo Fraidenraich$^{3}$\\
$^{1}$ School of Energy Systems, LUT University, Finland\\
$^{2}$ Centre Tecnol\`{o}gic de Telecomunicacions de Catalunya (CTTC/CERCA), Spain\\
$^3$ School of Electrical and Computer Engineering, University of Campinas, Brazil\\
e-mail: \{dick.carrillo.melgarejo, arthur.sena, pedro.nardelli\}@lut.fi, ckalalas@cttc.es,\\
gf@decom.fee.unicamp.br
}

\begin{document}

\maketitle

\thispagestyle{empty}
\pagestyle{empty}

\begin{abstract}
Reconfigurable intelligent surfaces (RISs) have been recently considered as one of the emerging technologies for future communication systems by leveraging the tuning capabilities of their reflecting elements. In this paper, we investigate the potential of an RIS-based architecture for uplink sensor data transmission in an ultra-reliable low-latency communication (URLLC) context. In particular, we propose an RIS-aided grant-free access scheme for an industrial control scenario, aiming to exploit diversity and achieve improved reliability performance. We consider two different resource allocation schemes for the uplink transmissions, i.e., dedicated and shared slot assignment, and three different receiver types, namely the zero-forcing, the minimum mean squared error (MMSE), and the MMSE-successive interference cancellation receivers.
Our extensive numerical evaluation
in terms of outage probability demonstrates the gains of our approach in terms of reliability, resource efficiency, and capacity and for different configurations of the RIS properties. An RIS-aided grant-free access scheme combined with advanced receivers is shown to be a well-suited option for uplink URLLC.

\end{abstract}

\begin{IEEEkeywords}
Reconfigurable intelligent surface, grant-free, URLLC, resource allocation.
\end{IEEEkeywords}

\begin{acronym}
  \acro{1G}{first generation of mobile network}
  \acro{1PPS}{1 pulse per second}
  \acro{2G}{second generation of mobile network}
  \acro{3G}{third generation of mobile network}
  \acro{4G}{fourth generation of mobile network}
  \acro{5G}{fifth generation}
  \acro{ARQ}{automatic repeat request}
  \acro{ASIP}{application specific integrated processors}
  \acro{AWGN}{additive white Gaussian noise}
   \acro{BER}{bit error rate}
  \acro{BCH}{Bose-Chaudhuri-Hocquenghem}
  \acro{BRICS}{Brazil-Russia-India-China-South Africa}
  \acro{BS}{base station}
  \acro{CDF}{cumulative density function}
  \acro{CoMP} {cooperative multi-point}
  \acro{CP}{cyclic prefix}
  \acro{CR}{cognitive radio}
  \acro{CS}{cyclic suffix}
  \acro{CSI}{channel state information}
  \acro{CSMA}{carrier sense multiple access}
  \acro{DFT}{discrete Fourier transform}
  \acro{DFT-s-OFDM}{DFT spread OFDM}
  \acro{DSA}{dynamic spectrum access}
  \acro{DVB}{digital video broadcast}
  \acro{DZT}{discrete Zak transform}
  \acro{eMBB}{enhanced mobile broadband}
  \acro{EPC}{evolved packet core}
  \acro{FBMC}{filterbank multicarrier}
  \acro{FDE}{frequency-domain equalization}
  \acro{FDMA}{frequency division multiple access}
  \acro{FD-OQAM-GFDM}{frequency-domain OQAM-GFDM}
  \acro{FEC}{forward error control}
  \acro{F-OFDM}{Filtered Orthogonal Frequency Division Multiplexing}
  \acro{FPGA}{field programmable gate array}
  \acro{FTN}{faster than Nyquist}
  \acro{FT}{Fourier transform}
  \acro{FSC}{frequency-selective channel}
  \acro{GFDM}{generalized frequency division multiplexing}
  \acro{GPS}{global positioning system}
  \acro{GS-GFDM}{guard-symbol GFDM}
  \acro{IARA}{Internet Access for Remote Areas}
  \acro{ICI}{intercarrier interference}
  \acro{IDFT}{Inverse Discrete Fourier Transform}
  \acro{IFI}{inter-frame interference}
  \acro{i.i.d.}{independent and identically distributed}
  \acro{IMS}{IP multimedia subsystem}
  \acro{IoT}{internet of things}
  \acro{IP}{Internet Protocol}
  \acro{ISI}{intersymbol interference}
  \acro{IUI}{inter-user interference}
  \acro{LDPC}{low-density parity check}
  \acro{LLR}{log-likelihood ratio}
  \acro{LMMSE}{linear minimum mean square error}
  \acro{LTE}{Long-Term Evolution}
  \acro{LTE-A}{Long-Term Evolution - Advanced}
  \acro{M2M}{Machine-to-Machine}
  \acro{MA}{multiple access}
  \acro{MAR}{mobile autonomous reporting}
  \acro{MF}{Matched filter}
  \acro{MIMO}{multiple-input multiple-output}
  \acro{MMSE}{minimum mean squared error}
  \acro{MRC}{maximum ratio combiner}
  \acro{MSE}{mean-squared error}
  \acro{MTC}{Machine-Type Communication}
  \acro{NEF}{noise enhancement factor}
  \acro{NFV}{network functions virtualization}
  \acro{OFDM}{orthogonal frequency division multiplexing}
  \acro{OOB}{out-of-band}
  \acro{OOBE}{out-of-band emission}
  \acro{OQAM}{offset quadrature amplitude modulation}
  \acro{PAPR}{peak-to-average power ratio}
  \acro{PDF}{probability density function}
  \acro{PHY}{physical layer}
  \acro{QAM}{quadrature amplitude modulation}
  \acro{PSD}{power spectrum density}
  \acro{QoE}{quality of experience}
  \acro{QoS}{quality of service}
  \acro{RC}{raised cosine}
  \acro{RRC}{root raised cosine}
  \acro{RTT} {round trip time}  
  \acro{SC}{single carrier}
  \acro{SC-FDE}{Single Carrier Frequency Domain Equalization}
  \acro{SC-FDMA}{Single Carrier Frequency Domain Multiple Access}
  \acro{SDN}{software-defined network}
  \acro{SDR}{software-defined radio}
  \acro{SDW}{software-defined waveform}
  \acro{SEP}{symbol error probability}
  \acro{SER}{symbol error rate}
  \acro{SIC}{successive interference cancellation}
  \acro{SINR}{signal-to-interference-and-noise ratio }
  \acro{SMS}{Short Message Service}
  \acro{SNR}{signal-to-noise ratio}
  \acro{STC}{space time code}
  \acro{STFT}{short-time Fourier transform}
  \acro{TD-OQAM-GFDM}{time-domain OQAM-GFDM}
  \acro{TTI}{time transmission interval}
  \acro{TR-STC}{Time-Reverse Space Time Coding}
  \acro{TR-STC-GFDMA}{TR-STC Generalized Frequency Division Multiple Access}
  \acro{TVC}{ime-variant channel}
  \acro{UFMC}{universal filtered multi-carrier}
  \acro{UF-OFDM}{Universal Filtered Orthogonal Frequency Multiplexing}
  \acro{UHF}{ultra high frequency}
  \acro{URLL}{Ultra Reliable Low Latency}
  \acro{V2V}{vehicle-to-vehicle}
  \acro{V-OFDM}{Vector OFDM}
  \acro{ZF}{zero-forcing}
  \acro{ZMCSC}{zero-mean circular symmetric complex Gaussian}
  \acro{W-GFDM}{windowed GFDM}
  \acro{WHT}{Walsh-Hadamard Transform}
  \acro{WLAN}{wireless Local Area Network}
  \acro{WLE}{widely linear equalizer}
  \acro{WLP}{wide linear processing}
  \acro{WRAN}{Wireless Regional Area Network}
  \acro{WSN}{wireless sensor networks}
  \acro{ROI}{return on investment}
  \acro{NR}{new radio}
  \acro{SAE}{system architecture evolution}
  \acro{E-UTRAN}{evolved UTRAN}
  \acro{3GPP}{3rd Generation Partnership Project }
  \acro{MME}{mobility management entity}
  \acro{S-GW}{serving gateway}
  \acro{P-GW}{packet-data network gateway}
  \acro{eNodeB}{evolved NodeB}
  \acro{UE}{user equipment}
  \acro{DL}{downlink}
  \acro{UL}{uplink}
  \acro{LSM}{link-to-system mapping}
  \acro{PDSCH}{physical downlink shared channel}
  \acro{TB}{transport block}
  \acro{MCS}{modulation code scheme}
  \acro{ECR}{effective code rate}
  \acro{BLER}{block error rate}
  \acro{CCI}{co-channel interference}
  \acro{OFDMA}{orthogonal frequency-division multiple access}
  \acro{LOS}{line-of-sight}
  \acro{VHF}{very high frequency}
  \acro{pdf}{probability density function}
  \acro{ns-3}{Network simulator 3}
  \acro{Mbps}{mega bits per second}
  \acro{EH}{energy harvesting}
  \acro{SWIPT}{simultaneous wireless information and power transfer}
  \acro{AF}{amplify-and-forward}
  \acro{DF}{decode-and-forward}
  \acro{WIT}{wireless information transfer}
  \acro{WPT}{wireless power transfer}
  \acro{FSFC}{frequency selective fading channel}
  \acro{DC}{direct current}
  \acro{FFT}{fast Fourier transform}
  \acro{RF}{radio frequency}
  \acro{SISO}{single-input single-output}
  \acro{RRC}{root raised cosine}
  \acro{TSR}{time-switching relaying}
  \acro{IFFT}{inverse fast Fourier transform}
  \acro{LIS}{large intelligent surfaces}
  \acro{URLLC}{ultra-reliable low-latency communication}
  \acro{ZMCSCG}{zero mean circularly symmetric complex Gaussian}
  \acro{PPSINR}{post-processing SINR}
  \acro{mMTC}{massive machine-type communication}
  \acro{NR}{New radio}
  \acro{RIS}{reconfigurable intelligent surface} 
  \acro{RAN}{radio access network}
  \acro{i.i.d.}{independent and identically distributed}
\end{acronym}

\section{Introduction}
\label{sec:introduction}
The \ac{5G} wireless systems aim to efficiently support three key generic services with broadly diverging operational requirements, i.e., \ac{eMBB}, \ac{mMTC}, and \ac{URLLC} \cite{5G_ref1, 5G_ref2, 5G_ref3, 5G_ref4}.
In particular, \ac{eMBB} services are associated with high data rate requirements with moderate reliability demands \cite{5G_ref5};
\ac{mMTC} typically involves a massive number of sporadically active devices transmitting delay-tolerant traffic in small data payloads \cite{5G_ref_mMTC};
and \ac{URLLC} entails low-latency transmissions of small payloads with very high reliability requirements often imposed by mission-critical application scenarios \cite{5G_ref_URLLC}.

In the particular case of the \ac{URLLC} service category, a wide range of emerging use cases are expected to be supported. Example application scenarios include
wireless control and automation in industrial factory environments, power system protection, inter-vehicular communications for improved traffic safety and efficiency, remote health services and the tactile Internet.
All these use cases require extremely low latency values (i.e., user-plane
radio latency $1$ms) and, simultaneously, 
an outage probability of less than $10^{-5}$ (i.e., reliability higher than $0.99999$) in terms of \ac{BLER}
\cite{5G_ref5}.
%
%
These stringent \ac{URLLC} requirements challenge the classical design principles of cellular networks and call for radical \ac{RAN} technical enablers.
%

Among the various technical enablers proposed for \ac{URLLC} enhancement, e.g., flexible numerology, mini-slot frame structure, link adaptation, etc., grant-free access protocols aim to minimize the latency associated with the uplink connection establishment. In grant-free multiple access, the signaling overhead can be significantly reduced by allowing \ac{UE}s to directly transmit their uplink data without sending scheduling requests to the \ac{BS} and waiting for uplink grant allocation.
However, the low protocol overhead comes at the cost of increased collisions among the contending \ac{UE}s, which may compromise the required reliability levels. 

Consequently, numerous variants of grant-free access protocols have been developed over the recent years. Solutions based on central coordination and feedback capabilities are rendered insufficient for mission-critical applications, especially at high-frequency bands and in harsh propagation environments where shadowing/blockage can cause direct links to the \ac{BS} to be in outage. An interesting approach relies on the \textit{a priori} allocation of the radio resources for each \ac{UE} transmission and subsequent retransmissions.
In \cite{kotaba_sharediversity}, the authors propose a resource-efficient transmission scheme where resources are shared among \ac{UE}s in a coordinated manner. Their approach improves reliability while relying on sophisticated receiver types, such as
\ac{MMSE}-\ac{SIC}. 

In this paper, we leverage the emerging paradigm of \ac{RIS} to enhance the performance of grant-free access. An \ac{RIS} consists of a number of programmable nearly passive elements with ultra-low power consumption, each of which can be properly tuned to apply arbitrary phase shifts to the impinging radio signals \cite{ntontin2019reconfigurable}. The potential of \ac{RIS}s has recently attracted extensive research attention and their applicability has been explored in various network scenarios, such as energy-efficient beamforming, physical-layer security, wireless power transfer, indoor positioning \cite{LIS_application_positioning ,LIS_application_beamforming,LIS_application_security}. Based on their unique capability of controlling the wireless propagation environment \cite{LIS_shlezinger,LIS_han,LIS_alouini}, \ac{RIS}s constitute a promising and low-cost solution to provide link diversity and achieve ultra-reliable uplink connectivity.  

In this context, we propose an \ac{RIS}-aided grant-free access scheme tailored for mission-critical \ac{URLLC} applications. In 
particular, we consider an \ac{RIS}-based network architecture for uplink data transmissions, aiming to exploit the additional degree of freedom for system optimization. Based on the benchmark framework followed in \cite{kotaba_sharediversity}, we consider two different uplink resource allocation schemes, i.e., dedicated and shared, for the \ac{UE} transmissions and we evaluate the performance in terms of outage probability for different receiver types. Our numerical simulations reveal the reliability and capacity gains of the \ac{RIS}-aided grant-free access compared to the performance of the legacy grant-free access method. Besides allowing for reduced receiver complexity, our proposed scheme improves the resource utilization as it requires less number of slots for \ac{UE} retransmissions. Finally, in an effort to demonstrate the \ac{RIS} capability in controlling the radio environment, we quantify the impact of the main \ac{RIS} features, i.e., number of elements and phase shifting values, in the overall performance.

%
%

\textit{Organization}: The rest of this paper is organized as follows.
We present the system model in Section \ref{sec:system_model}.
In Section \ref{sec:performance_analysis}, we provide the characteristics of the considered receiver types and we define the outage probability, which constitutes the key performance metric of our analysis.
%
In Section \ref{sec:results}, we present and discuss the numerical results obtained
through extensive simulations. Finally, concluding remarks and future research directions are drawn in Section \ref{sec:conclusions}.

\textit{Notation}: The notation used in this paper is the following: $(\cdot)^H$ denotes the conjugate transpose, $(\cdot)_{i,j}$ represents the $(i,j)^{th}$ entry of a matrix, 
the uppercase and lowercase boldface letters denote matrices and vectors respectively, while $\matriz{I}_\text{N}$ represents an $N \times N$ identity matrix.



\section{System Model}
\label{sec:system_model}

In this section, we present the signal model for the uplink grant-free transmission assisted by an \ac{RIS}. 

\begin{figure}[t!]
   \centering
   \includegraphics[width=0.8\textwidth]{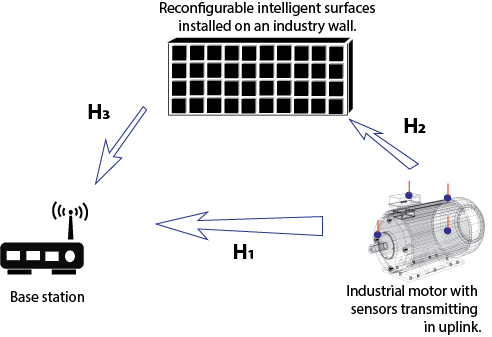}
   \caption{System model of an \ac{RIS}-aided uplink sensor data transmission in a wireless industrial environment. The \ac{RIS} is deployed on a nearby wall in the vicinity of both communication ends.}
   \label{fig:systemmodel}
 \end{figure}
 
As illustrated in Fig. \ref{fig:systemmodel}, we consider a wireless industrial control scenario where a single \ac{BS} is serving $N$ \ac{URLLC}-type sensors attached to a motor to support a control-loop. The sensors transmit their mission-critical data related to the motor states in periodic uplink frames composed of $M$ slots, each of which consists of $L$ channel uses. Without loss of generality, it is assumed that the packet length of each sensor occupies $1$ slot. We further consider a Rayleigh fading model for the channel, which remains constant over all $L$ uses of the slot. The received signal, $\matriz{Y} \in \mathds{C}^{M\times L}$, at the \ac{BS} can be written as
\begin{equation}
\label{eq:general_model}
    \matriz{Y} = \matriz{H} \matriz{X} + \matriz{N},
\end{equation}
where $\matriz{X} \in \mathds{C}^{N\times L}$ contains the complex modulated symbols of the sensors with transmit power $E[|x_{i,j}|^2]=P_x$; $\matriz{H} \in \mathds{C}^{M\times N}$ 
contains the channel gains of the sensors in the available slots;
and $\matriz{N} \in \mathds{C}^{M\times L}$ is an additive white Gaussian noise with zero mean and variance $\sigma^2$. We further assume that the total transmitted power per sensor is independent of the number of transmissions.

Besides the direct signal path between the sensors and the \ac{BS}, communication also takes place via the \ac{RIS} which consists of $K$ passive reflecting elements deployed in the vicinity of both communication ends. Thus, $\matriz{H}$ can be decomposed as
\begin{equation}
    \matriz{H} = \matriz{H}_1 + \matriz{H}_3  \pmb{\Phi} \matriz{H}_2,
\end{equation}
where $\matriz{H}_1 \in \mathds{C}^{M\times N}$ represents the direct channel between the sensors and the \ac{BS}; $\matriz{H}_3 \in \mathds{C}^{M\times K}$ corresponds to the channel between the \ac{RIS} and the \ac{BS}; $\pmb{\Phi}  \triangleq \text{diag}[\phi_1, \phi_2, \dots,\phi_K]$ is a diagonal matrix accounting for the effective phase shifting values $\phi_k, \ \forall k = 1, 2, \dots , K$, applied by all \ac{RIS} reflecting elements; and $\matriz{H}_2 \in \mathds{C}^{K\times N}$ represents the channel between the sensors and the \ac{RIS}.
The entries of $\matriz{H}_1, \matriz{H}_2 , \matriz{H}_3$ are also modeled as \ac{i.i.d.} \ac{ZMCSCG} variables. The effective phase shifting value for the $k$-th element of the \ac{RIS} is defined as 
\begin{equation}
    \label{eq:lis_phasesifting}
    \phi_k \in \mathcal{F} \triangleq \bigg\{  e^{\frac{j2\pi m}{2^b}} \bigg\}_{m=0}^{2^b-1}\enskip,
\end{equation}
where $\mathcal{F}$ is the set with the available phase shifting values; $j \triangleq \sqrt{-1}$ is the imaginary unit; $m$ represents the phase shifting index; and $b$ is the phase resolution in number of bits \cite{LIS_chongwen}. The number of phase shifting values per \ac{RIS} element is $2^b$. 

%
Two different resource allocation schemes for the uplink grant-free transmissions of the sensors\footnote{Similar to \cite{kotaba_sharediversity}, 
we assume that an initial uplink parameter configuration has been already performed during the network registration of the sensors. This can be carried out as part of the system information broadcasted by the \ac{BS} over the cell.} are considered at the \ac{BS}. In particular,
\begin{enumerate}
    \item A \textit{dedicated} resource allocation scheme, where every sensor transmission is assigned a distinct slot. This scheme constitutes a robust approach against mutual interference by allocating orthogonal resources to the sensors. However, despite the controlled interference environment, this scheme may prove inefficient in terms of resource utilization.
    \item A \textit{shared} resource allocation scheme, where all sensors are able to share the available slots among them. In stark contrast to the dedicated assignment, this scheme is more spectrum-efficient at the expense of induced intra-slot interference in the sensors' transmissions.
\end{enumerate}

\section{Performance Analysis}
\label{sec:performance_analysis}
Based on the system model defined in Section \ref{sec:system_model}, we consider a generic \ac{RIS}-based \ac{MIMO} framework for the purpose of the analysis, where
each sensor corresponds to a single
transmit antenna, and each time-frequency slot is served by
a different virtual receive antenna
\cite{kotaba_sharediversity}.
This approach allows us to assess the performance using standard tools 
originally derived for a \ac{MIMO} system analysis. 

The estimation of the original transmitted signal is contaminated by the inter-stream interference from the other sensors and the noise. Assuming perfect channel estimation, the detected signal can be expressed as
\begin{equation}
\label{eq:detectionmodel}
    \hat{\matriz{X}} = \matriz{F}\matriz{Y} = \matriz{F}\matriz{H}\matriz{X}+\matriz{F}\matriz{N},
\end{equation}
where $\matriz{F}$ depends on the receiver design. In what follows, we consider three different  types of \ac{MIMO} receivers, i.e.,
\ac{ZF}, \ac{MMSE}, and \ac{MMSE}-\ac{SIC}, for the recovery of the transmitted symbols of the sensors \cite{MIMO_book}. In particular,
%
\begin{enumerate}
    \item A \ac{ZF} receiver which focuses on completely eliminating the
\ac{ISI} with no control over the energy in the stream of interest. In particular, it aims to minimize the mean squared estimation error (averaged over the
noise) on the symbols under the constraint of complete elimination of ISI.
To estimate the received signal of the form defined in (\ref{eq:general_model}), the receiver applies the following \ac{ZF} detection matrix $\matriz{F}_\text{zf}$ given by
\begin{equation}
    \label{detec_matrix_zf}
    \matriz{F}_\text{zf} = \matriz{H} \ (\matriz{H}^H \matriz{H})^{-1} \matriz{H}^H.
\end{equation}

\item
An \ac{MMSE} receiver which aims to minimize the average estimation error on the
transmitted symbols and  the average is taken over the transmitted symbols and the noise. Different from the \ac{ZF} receiver, the \ac{MMSE} does not require complete \ac{ISI} elimination. The equivalent expression of the \ac{MMSE} receiver, $\matriz{F}_\text{mmse}$, used to estimate the received signal is given by

\begin{equation}
    \label{detec_matrix_mmse}
    \matriz{F}_\text{mmse} =  \bigg(\matriz{H}^H \matriz{H} + \frac{\sigma^2}{P_x} \matriz{I}\bigg)^{-1} \matriz{H}^H.
\end{equation}

\item
An \ac{MMSE}-\ac{SIC} receiver which relies on an iterative principle where, at each iteration, a data stream is decoded considering the other streams as interference. The decoded stream is then removed from the list of interfering streams and, at the next iteration, the selected stream is decoded with one less interferer.
The detection matrix for the \ac{MMSE}-\ac{SIC} receiver has the same expression as in \eqref{detec_matrix_mmse}.
However, the procedure to calculate the \ac{PPSINR} is iterative with optimal ordering. An \ac{MMSE}-\ac{SIC} receiver design improves the detection accuracy at the cost of increased computational complexity compared to the previous receivers.
\end{enumerate}
%

%
Based on Eq. (\ref{eq:detectionmodel}), the \ac{PPSINR} of a sensor data stream $i$ can be expressed as
\begin{equation}
\label{eq:ppsinr}
    \text{SINR}(i) = \frac{P_x L |(\matriz{F}\matriz{H})_{i,i}|^2}{P_x L \sum_{j\neq i} |(\matriz{F}\matriz{H})_{i,j}|^2 + (E[\matriz{N}\matriz{N}^H])_{i,i}}. 
\end{equation}
For the performance assessment of 
mission-critical \ac{URLLC} scenarios, as the one considered in this paper, the outage probability per sensor data stream is chosen as the most appropriate reliability metric. In general, the outage probability constitutes a widely-used information theoretic tool to evaluate the performance in MIMO systems. For a selected transmission rate $R$, it is computed as 
\begin{equation}
    \label{outage}
    p_\text{out}(i) = Pr\{R > \mathcal{R}_\text{max}(i)\},
\end{equation}
where $\mathcal{R}_\text{max}(i) = \text{log}_2(1+\text{SINR}(i))$ corresponds to the maximal achievable rate for reliable communication of sensor data stream $i$.
Using \eqref{outage}, the 
outage probability for each receiver design can be computed individually.
%

\section{Numerical Results}
\label{sec:results}
As described in Section \ref{sec:performance_analysis}, we 
aim to assess the performance from the viewpoint of 
a \ac{MIMO} framework.
%
An \ac{RIS}-aided $i$-sensor \ac{MIMO} communication model is therefore considered, as illustrated in Fig. \ref{fig:systemmodel}. The reliability performance of grant-free access is evaluated in terms of outage probability for different receiver types, i.e., \ac{ZF}, \ac{MMSE}, and \ac{MMSE}-\ac{SIC}, and uplink resource allocation schemes, i.e., dedicated and shared. The sensors are equipped with omni-directional antennas and transmit their data simultaneously. The \ac{RIS} elements are deployed uniformly on an ideal rectangular surface and the effective phase shifting values defined in Eq. \eqref{eq:lis_phasesifting} are considered to have low phase resolution tuning capabilities, i.e., $b=1$.

In order to obtain an averaged system performance, the achievable rates for each of the considered receiver types are averaged over $2 \cdot 10^5 $ channel realizations drawn from an \ac{i.i.d.} Rayleigh distribution.
%
%
A sensor transmission rate $R=2$ bits/s/Hz is also assumed for the outage probability calculations.

\begin{figure}[t!]
   \centering
   \includegraphics[width=1\textwidth]{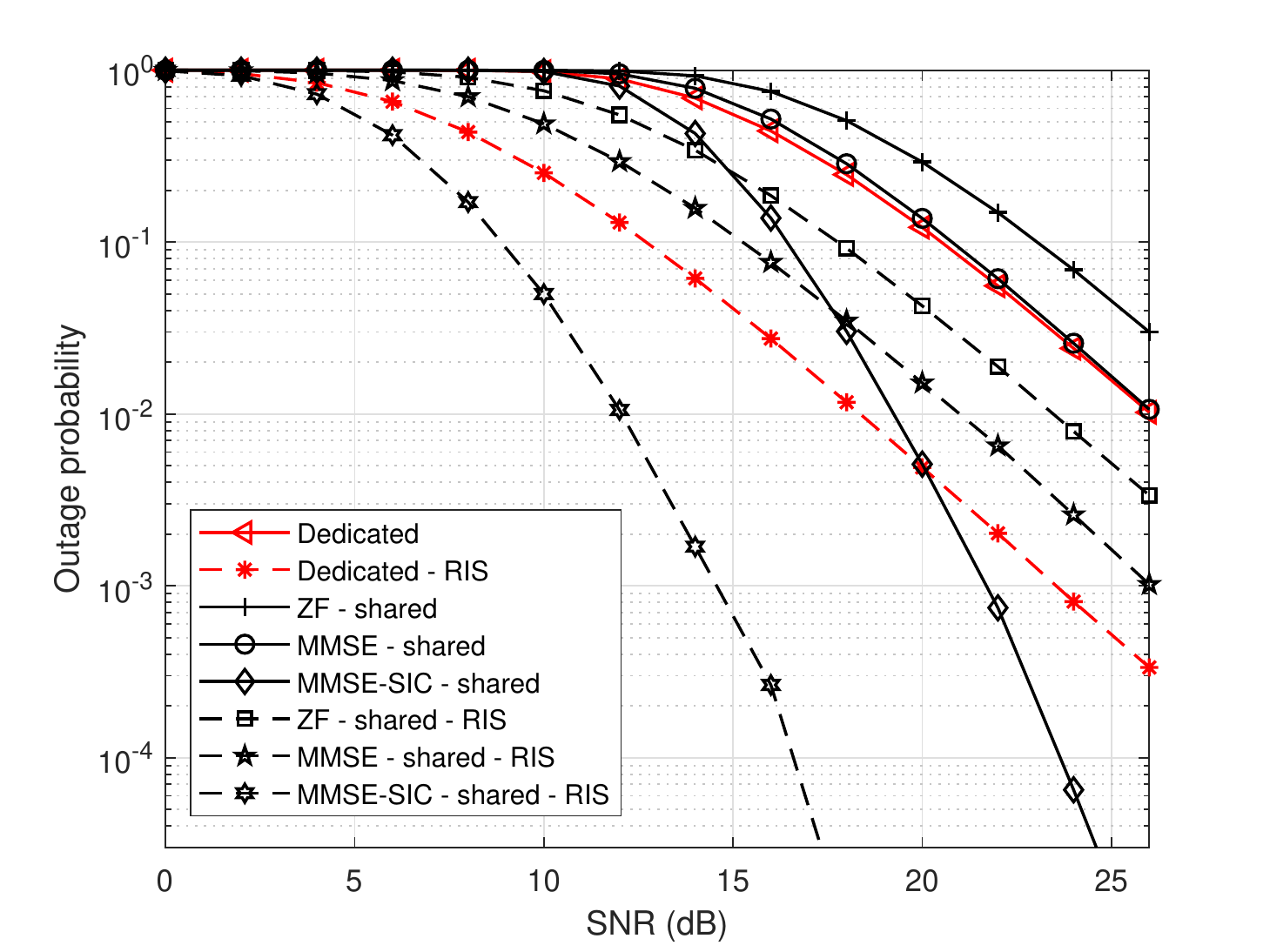}
   \caption{Performance comparison of the different receivers and resource allocation schemes for a grant-free transmission of $N=5$ sensors with/without the assistance of an \ac{RIS} with $K=6$ elements. A number of $M=11$ slots is considered for the dedicated scheme whereas $M=6$ for the shared one.}
   \label{fig:withoutLIS}
 \end{figure}

Fig. \ref{fig:withoutLIS} illustrates the performance comparison of the different receiver types and resource allocation schemes for a grant-free transmission with/without the assistance of an \ac{RIS}. 
We can observe that a grant-free scheme with shared slots and a \ac{ZF} receiver leads to a degraded performance with respect to the grant-free scheme with dedicated slot assignment. However, as the receiver complexity increases, we note that the shared grant-free scheme gradually achieves superior performance compared to the dedicated case. In addition, this reliability gain is attained with relatively less (nearly half) slots, which reveals the resource efficiency of the shared approach. In both resource allocation methods, the reliability gains of an \ac{RIS}-aided grant-free access can be easily identified for all receiver types. Interestingly, we can observe that an \ac{MMSE} receiver in an \ac{RIS}-aided system outperforms the sophisticated \ac{MMSE}-\ac{SIC} receiver in a system with no \ac{RIS} for \ac{SNR} values until $18$dB. Thus, the presence of \ac{RIS} may allow the use of receivers with lower computational complexity to achieve acceptable outage performance. Finally, we remark that the use of \ac{RIS} in combination with advanced receiver designs can achieve significantly low outage probability values, revealing their potential in supporting \ac{URLLC}.

\begin{figure}[t!]
   \centering
   \includegraphics[width=1\textwidth]{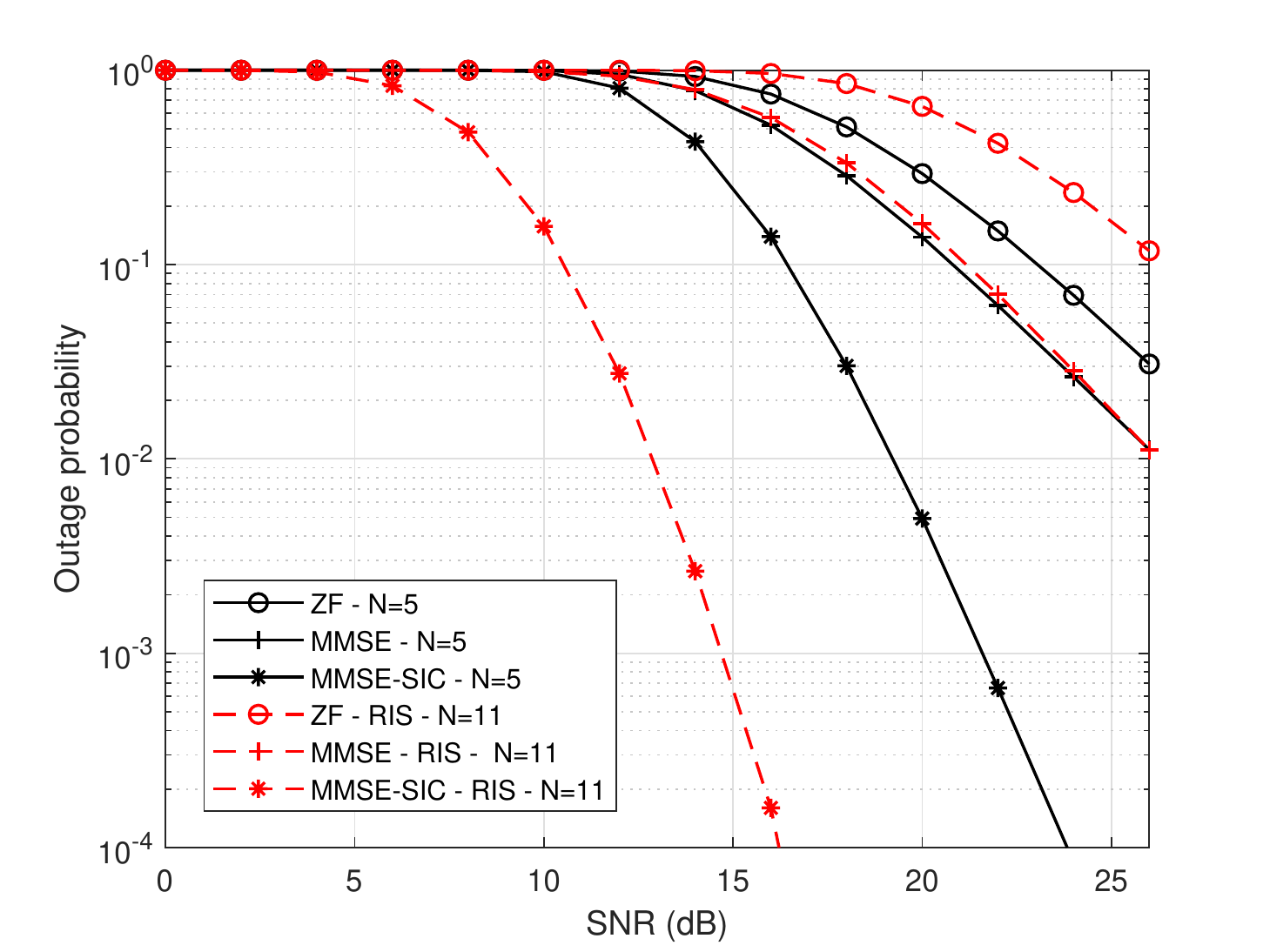}
   \caption{Performance comparison of a non-\ac{RIS} grant-free access scheme supporting $N=5$ sensors and an \ac{RIS}-aided grant-free access scheme supporting $N=11$ sensors. A shared resource allocation strategy is considered and the number of \ac{RIS} elements is $K=6$.}

   \label{fig:withLIS1}
 \end{figure}

Fig. \ref{fig:withLIS1} corroborates 
the capacity gains achieved with an \ac{RIS}-aided grant-free access scheme. In particular, an \ac{MMSE} receiver in an RIS-aided system supporting $11$ sensors simultaneously transmitting their data, achieves nearly similar outage performance compared to an \ac{MMSE} receiver in a non-\ac{RIS} setup with $5$ sensors. Thus, an \ac{RIS}-aided grant-free scheme is able to support more than double of the number of sensors using the same amount of slots. The capacity potential of \ac{RIS} is further enhanced when a higher-complexity receiver, i.e., \ac{MMSE}-\ac{SIC}, is considered.

\begin{figure}[!h]
   \centering
   \includegraphics[width=1\textwidth]{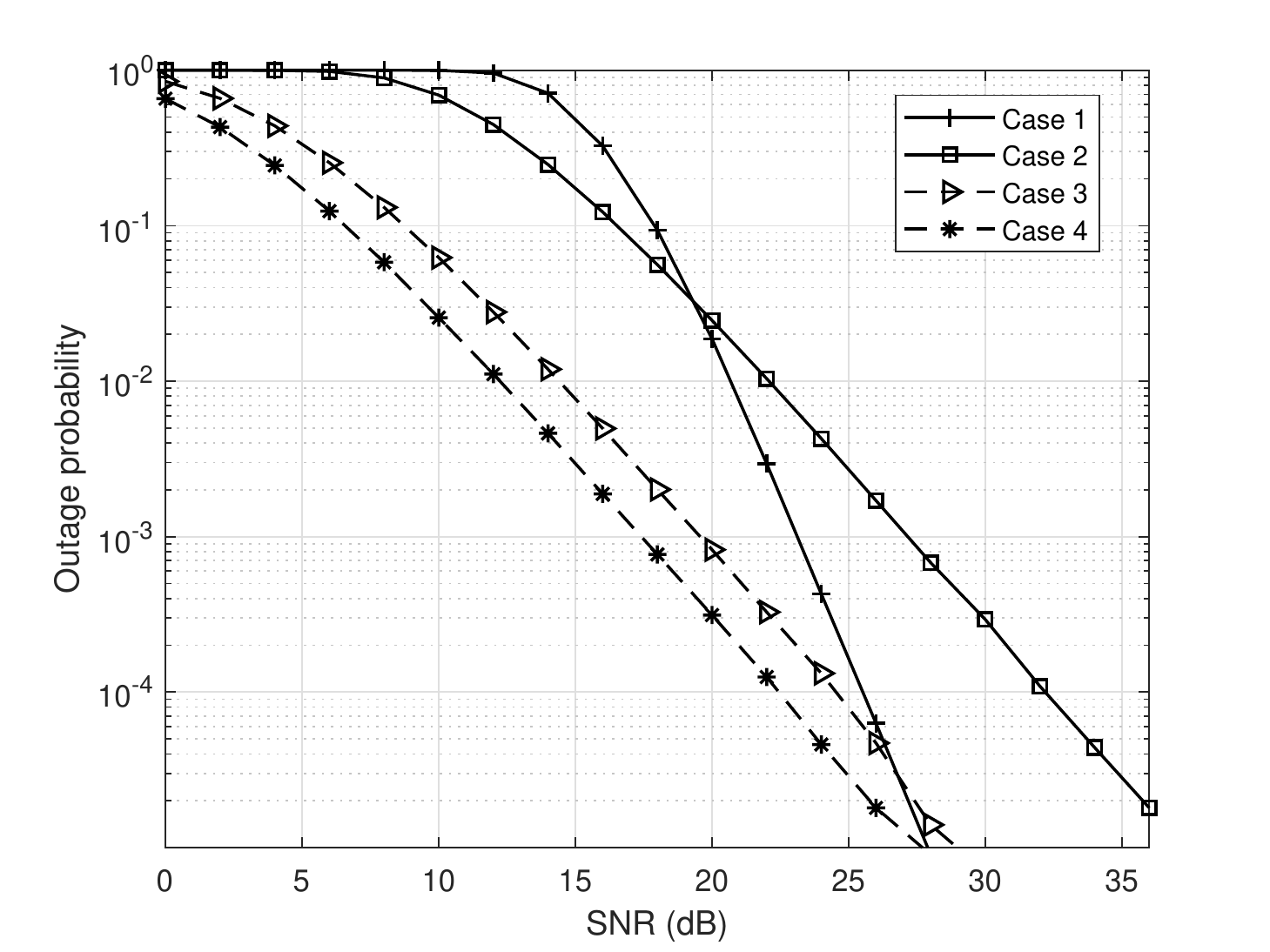}
   \caption{Performance comparison of different retransmission policies for a dedicated resource allocation scenario with $N=2$ sensors and $M=10$ slots. The following cases are illustrated: Case 1: $4$ retransmission slots per sensor with no-\ac{RIS} assistance; Case 2: 1 retransmission slot per sensor with no-\ac{RIS} assistance; Case 3: 1 retransmission slot per sensor with an \ac{RIS} of $K=6$ elements; Case 4: 1 retransmission slot per sensor with an \ac{RIS} of $K=8$ elements.}
   \label{fig:withLIS2}
 \end{figure}

Fig. \ref{fig:withLIS2} illustrates how the different retransmission strategies compare in terms of outage probability for a grant-free scenario of dedicated slot allocation with/without the assistance of an \ac{RIS}. In particular, each of the $N=2$ sensors is assigned a dedicated slot for its data transmission while two allocation cases are considered for their retransmissions: \textit{i}) 4 dedicated slots per sensor; and \textit{ii}) 1 dedicated slot per sensor. While for a non-\ac{RIS} system, the first case (i.e., 4 slots/sensor) is more robust in the high \ac{SNR} regime compared to the more conservative second case (i.e., 1 slot/sensor), we can observe that an \ac{RIS}-aided system can significantly improve the performance even for a smaller amount of retransmission slots. It can be also noticed that the outage probability further decreases when the number of $K$ elements of the \ac{RIS} increases.

 %
Finally, Table I shows the impact of the tuning capabilities of the \ac{RIS} elements on the outage probability performance for a grant-free scenario of shared resources with an \ac{MMSE}-\ac{SIC} receiver. In particular, we consider an \ac{RIS} consisting of $K=3$ elements with 1-bit phase resolution, i.e., $\phi_k=1$ when $m=0$, and $\phi_k=-1$ when $m=1$, for $k=1,2,3$. As it can be observed, for a simulation of $5 \cdot 10^6$ channel realizations, a proper configuration of the phase shifting values among the \ac{RIS} elements affects the outage performance. These findings corroborate 
the flexibility offered by an \ac{RIS}-based architecture, even with low phase resolution reflecting elements, in \ac{URLLC} scenarios with stringent outage requirements, e.g., in the order of $10^{-5}$.

\begin{table}[t!]
\begin{center}
 \label{tab:phi_values}
 \caption{Performance comparison of the different combinations of phase shifting values for 1-bit phase resolution ($b=1$), $R=1.5$ bits/s/Hz and SNR = $24 \mathrm{dB}$.}
 \begin{tabular}{ c | c | c | c |c} 
\hline 
\textbf{Combination} & \textbf{Element 1} & \textbf{Element 2} & \textbf{Element 3} & \textbf{Outage}\\ [0.5ex] 
 \hline\hline
 1 & 1 & 1 & 1 & $1.1  \cdot 10^{-5}$\\  
 \hline
 2 & 1 & 1 & -1& $0.96  \cdot 10^{-5}$ \\
 \hline
 3 & 1 & -1 & 1& $1.18  \cdot 10^{-5}$ \\
 \hline
 4 & 1 & -1 & -1& $1.3  \cdot 10^{-5}$ \\
  \hline
 5 & -1 & 1 & 1& $1.04  \cdot 10^{-5}$ \\
  \hline
 6 & -1 & 1 & -1& $0.9  \cdot 10^{-5}$ \\
  \hline
 7 & -1 & -1 & 1& $1.0  \cdot 10^{-5}$ \\
 \hline
 8 & -1 & -1 & -1& $1.02  \cdot 10^{-5}$ \\ 
\hline
\end{tabular}
\end{center}
\end{table}

\section{Conclusions}
\label{sec:conclusions}
In this paper, we propose a novel grant-free access scheme assisted by an \ac{RIS} architecture and tailored for uplink \ac{URLLC} use cases. We consider two different resource allocation schemes for the uplink data transmissions, i.e., dedicated and shared slot assignment, and we evaluate the performance in terms of outage probability for linear receivers of different computational complexity. Our extensive simulation results showcased that an \ac{RIS}-aided grant-free access scheme in combination with advanced \ac{SIC} receivers substantially improves the reliability performance, even at the low \ac{SNR} regime. In addition, we 
quantified the resource utilization and capacity gains offered by our proposed scheme. Finally, we showed the impact of the tuning capabilities of an \ac{RIS}, i.e., different number of elements and phase shifting values, on the outage probability. The overall performance assessment revealed the suitability of an \ac{RIS}-based framework for the support of mission-critical uplink \ac{URLLC} scenarios associated with stringent requirements.

Future interesting lines of research include the consideration of energy efficiency aspects for the \ac{RIS}-aided grant-free access as well as the joint design of the transmit sensor powers and the phase shifting values for the \ac{RIS} elements. An extension of this work would involve the study of the impact of channel estimation and feedback overhead on the performance of the \ac{RIS}-aided grant-free system.  Other impairment to be evaluated in further studies is related to the delay path impact between sensors and base station and sensors and \ac{RIS} elements, specially in the \ac{URLLC} scenario.

\section*{Acknowledgements}
This work is partly funded by Academy of Finland via ee-IoT project (ICT2023/n.319009), via FIREMAN consortium (CHIST-ERA/n.326270), and via EnergyNet (Research Fellow/n.321265\&328869), and by the Spanish National Foundation (PCI2019-103780), and the Generalitat de Catalunya under Grant 2017 SGR 891.


\bibliographystyle{IEEEtran}
\bibliography{references.bib}

\end{document}